\documentclass[conference]{IEEEtran}
\IEEEoverridecommandlockouts

\usepackage[usenames,dvipsnames,svgnames,table]{xcolor}
\usepackage{graphicx, float, wrapfig, caption, subcaption}
\usepackage[colorlinks=true, pdfstartview=FitV, linkcolor=Emerald, citecolor=cyan, urlcolor=WildStrawberry]{hyperref}

\usepackage{algorithmic}
\usepackage{textcomp}
\def\BibTeX{{\rm B\kern-.05em{\sc i\kern-.025em b}\kern-.08em
    T\kern-.1667em\lower.7ex\hbox{E}\kern-.125emX}}

\newcommand{\orcid}[1]{%
  \href{https://orcid.org/#1}{\includegraphics[height=1.7ex]{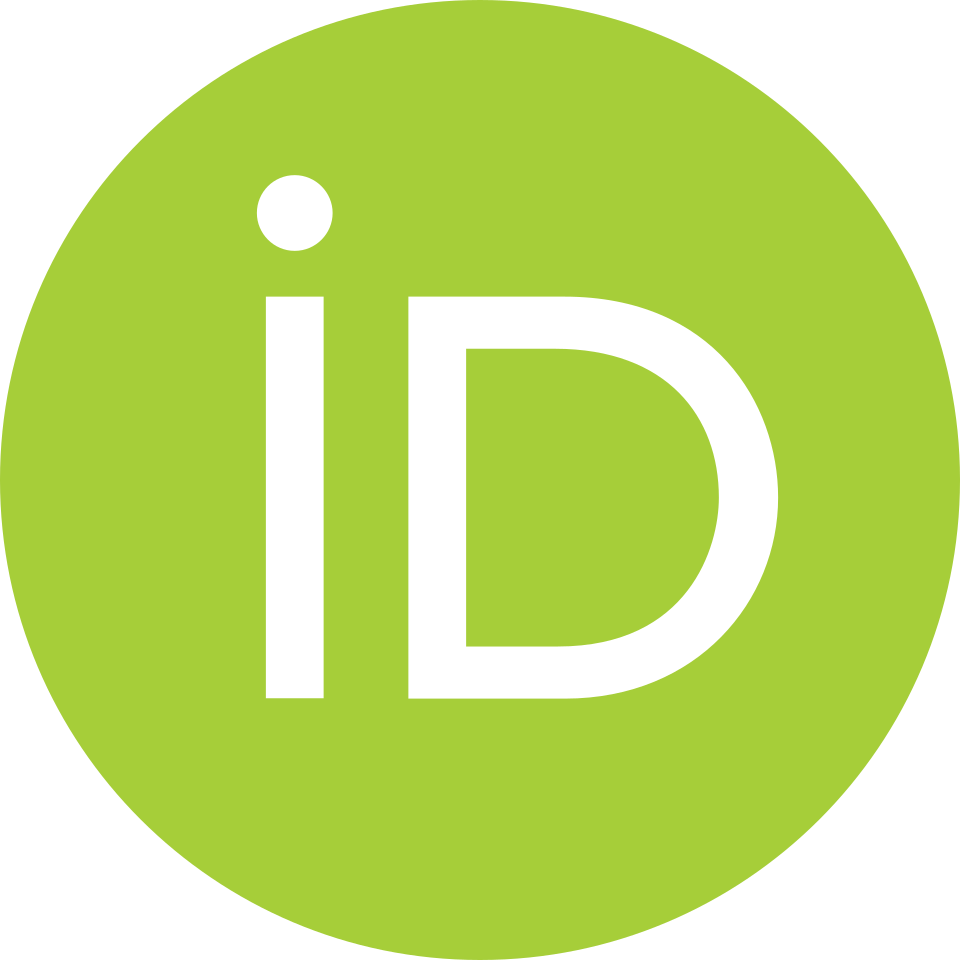}}%
}

\newcommand{\statementcolor}{black}

\usepackage[IEEEFUSION]{beesonmacros}

\newcommand{\mN}{\mathcal{N}}

\newcommand{\wt}[1]{\widetilde{#1}}

\newcommand{\mF}{\mathcal{F}}

\usepackage[acronyms, nopostdot, nonumberlist, nogroupskip]{glossaries}

\makeglossaries

\newcommand*{\Acro}[4][]{%
\ifthenelse { \equal {#1} {} }%
{ \newacronym{#2}{#3}{#4} }%
{ \newacronym[#1]{#2}{#3}{#4} }%
}

\Acro{PF}{PF}{Particle Filter}
\Acro{nPF}{nPF}{Nudged Particle Filter}
\Acro{Var-nPF}{Var-nPF}{Variational nudged Particle Filter}
\Acro{KF}{KF}{Kalman Filter}
\Acro{EnKF}{EnKF}{Ensemble Kalman Filter}
\Acro{UKF}{UKF}{Unscented Kalman Filter}
\Acro{GMM}{GMM}{Gaussian Mixture Model}
\Acro{FPF}{FPF}{feedback particle filter}
\Acro{MAP}{MAP}{maximum a posteriori}
\Acro{BM}{BM}{Brownian motion}
\Acro{PDE}{PDE}{partial differential equation}
\Acro{RMSE}{RMSE}{Root Mean Square Error}
\Acro{ESS}{ESS}{Effective Sample Size}
\Acro{nESS}{nESS}{normalized ESS}
\Acro{sc4DVar}{sc4DVar}{strong constraint 4-dimensional variational}
\Acro{OCP}{OCP}{optimal control problem}
\Acro{MC}{MC}{Monte Carlo}
\Acro{L63}{L63}{Lorenz 1963}
\Acro{TU}{TU}{time units}
\Acro{VF}{VF}{vector field}

\begin{document}

\title{
A Variational Pseudo-Observation Guided \\
Nudged Particle Filter
}

\author{
\IEEEauthorblockN{Theofania Karampela and Ryne Beeson\,\orcid{0000-0003-2176-0976}} 
\IEEEauthorblockA{\textit{Mechanical and Aerospace Engineering} \\
\textit{Princeton University}\\
Princeton, New Jersey, USA \\
tk4161@princeton.edu and ryne@princeton.edu}
}

\maketitle

\thispagestyle{plain}
\pagestyle{plain}

\begin{abstract}
Nonlinear filtering with standard \gls*{PF} methods requires mitigative techniques to quell weight degeneracy, such as resampling. 
This is especially true in high-dimensional systems with sparse observations.
Unfortunately, such techniques are also fragile when applied to systems with exceedingly rare events. 
Nonlinear systems with these properties can be assimilated effectively with a control-based \gls*{PF} method known as the \gls*{nPF}, but this method has a high computational cost burden. 
In this work, we aim to retain this strength of the nudged method while reducing the computational cost by introducing a variational method into the algorithm that acts as a continuous pseudo-observation path. 
By maintaining a \gls*{PF} representation, the resulting algorithm continues to capture an approximation of the filtering distribution, while reducing computational runtime and improving robustness to the ``rare" event of switching phases. 
Preliminary testing of the new approach is demonstrated on a stochastic variant of the nonlinear and chaotic \gls*{L63} model, which is used as a surrogate for mimicking ``rare" events. 
The new approach helps to overcome difficulties in applying the \gls*{nPF} for realistic problems and performs favorably with respect to a standard \gls*{PF} with a higher number of particles.
\end{abstract}

\vspace{0.25em}

\section{Introduction}
\label{section: introduction}

In this manuscript we are interested in the filtering solution to partially observable continuous-time signal processes $(X_t)$ with discrete-time observations $(Y_{t_k})$ 
\EquationAligned{
\label{equation: basic SDE filtering setup}
dX_t 
&= f(X_t) dt + \sigma(X_t) dW_t, &\quad X_0 &= x \in \R{m}, \\
Y_{t_k} 
&= h(X_{t_k}) + \xi_{t_k},  &\quad Y_0 &= 0 \in \R{d}, 
}
where $(t_k)$ is an increasing sequence of observation times indexed by $k \in\N{}$ with $t_0 = 0$, 
and $\xi_{t_k} \sim \mN(0, \Sigma_y)$ is the observation noise that is independent of the standard \gls*{BM} $W_t$ that drives the signal process $X_t$. 
We denote the filtering solution of $X$ at time $t$ given the observations up to time $t_k \leq t$ by $p_t(x | \mF^Y_k)$.
We use $\mF^Y_k$ to denote all observations up to time $t_k$ (simplifying the subscript from $t_k$ to just $k$ for brevity here and in what follows) and hence the prior conditional distribution at time $t_k$ is $p_k(x | \mF^Y_{k - 1})$ and the posterior is $p_k(x | \mF^Y_k)$. 

Additional properties that we assume of \eqref{equation: basic SDE filtering setup}, and that drive choices in filtering algorithms and the method developed in this work, is that $X$ represents a state of very high dimension (e.g., in the extreme cases of geophysical data assimilation \cite{vanLeeuwen:2019qjrms,Synder2008}, this can be $m \sim O(10^8)$ degrees of freedom that results from the discretization of \glspl*{PDE} describing physical processes), that the prior and posterior distributions are non-Gaussian, that the numerical solution of the signal process given in \eqref{equation: basic SDE filtering setup} is computationally demanding, and lastly, that intermittent aperiodic behavior characteristic of rare events takes place. 
A rare event is one that is represented by a low probability of occurrence; often associated with a state $X$ in the tail of the distribution $p(x | \mF^Y)$. 
Geophysical models generally possess the first three properties \cite{vanLeeuwen:2019qjrms, Evensen:2022.springer.international}, and examples of rare events include hurricanes, earthquakes, and rapid overturning in ocean flows \cite{Bucklew:2004.springer.textbook, Vanden-Eijnden:2012.cpam.65.12, Qiu:2000.jpo.30.8}. 
Filtering problems with these characteristics lead naturally to methods of either a variational or a low-sample ensemble-based approach. 

\begin{figure}[t]
\centering
\includegraphics[width=\columnwidth]{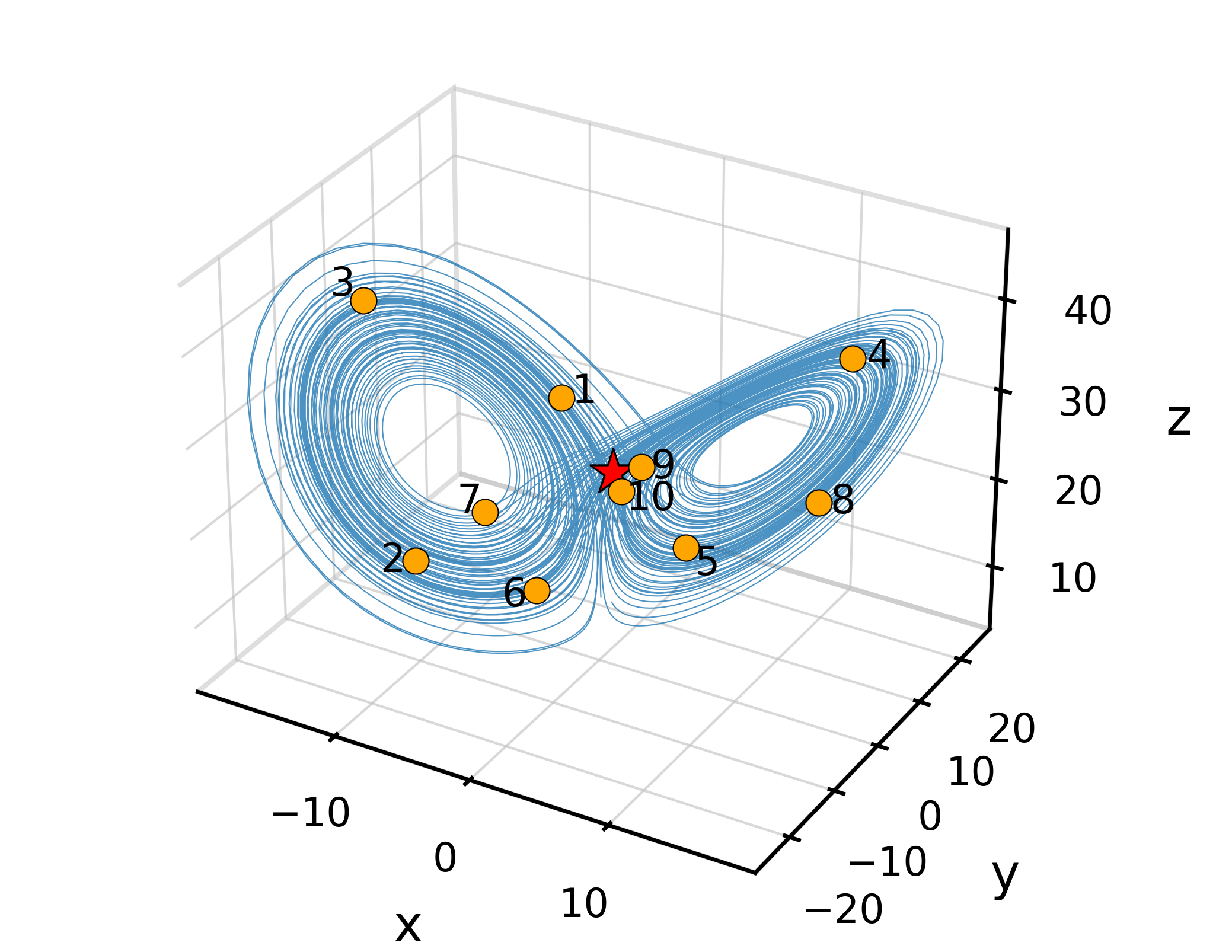}
\caption{
\gls*{L63} deterministic attractor with initial conditions (see  Table~\ref{tab:mc_ic_sweep_ess_rmse}) used in numerical experiments (see \S \ref{section: numerical experiments}). 
}
\label{fig:l63_ic_locations}
\end{figure}

Variational methods provide a smoothing solution for the \gls*{MAP} state \cite{Evensen:2022.springer.international, Talagrand:1987.qjrms.113.478, Courtier:1987.qjrms.113.478}. 
Their disadvantage is that they only provide a MAP state and not a representation of the full distribution or even higher-order moments. 
Lastly, because of the computational difficulty of solving such problems for high-dimensional systems, parameters that should be state and time varying are often fixed to static background (or climatological) values and hence do not represent the true near term variability. 

An alternative solution is to approximate the filtering distribution with an importance sampling approach. 
With $((X^i, w^i) \sim p)_{i = 1}^N$ representing $N \in \N{}$ pairs of samples $(X^i)$ with normalized weights $(w^i)$ from a distribution $p$ , the $N$-ensemble approximation $p^N$ to $p$ is
\begin{align}
\label{equation: ensemble approximation of a distribution}
p^N (dx)
\equiv \sum_{i = 1}^N w^i \delta_{X^i}(dx) 
\approx p (dx),
\end{align}
where $\delta_{X^i}$ is the Dirac distribution at $X^i$. 
The class of \gls*{PF} methods allow the weights to change during assimilation cycles; a reflection of the fact that they are importance weights. 
In the standard approach, the particle locations remain fixed during the Bayesian update. 
As we will recall in \S \ref{section: standard and nudged particle filter}, the adjustment of the particle weights at times of filtering updates is what ultimately leads to the well known issue of particle degeneracy. 
Some recent efforts in the literature navigate the issue of particle degeneracy by maintaining equal weights during Bayes' update by reframing the discrete-time update as a flowing of the particles in the state-space during a pseudo-observation interval \cite{Daum:2007.sdpst, Hanebeck:2003.isop.5099, Yang:2013.ieee.tac.58.10}, solving for an equivalent flow map \cite{Spantini:2022.siam.review.64.4, Reich:2019.an.28}, or an implicit mapping \cite{Chorin:2009.pnas.106.41}. 
Additionally, comments on how the flow-based approaches distinguish themselves from the main method of this paper, based on the \gls*{nPF}, will be given in \S \ref{section: standard and nudged particle filter}.

There are filtering methods that are ensemble-based, where the weights are chosen as equal (i.e., $w^i = 1 / N$) and remain equal throughout the filtering update. 
Common families of these methods include the well-known variants of the \gls*{EnKF} \cite{Evensen:1994.jgr.o.99.c5, Burgers:1998.mwr.126.6}, \gls*{UKF} \cite{Julier:2004ieee92.3}, and \gls*{GMM} \cite{Sorenson:1971auto.7.4}.
Although these methods provide a built-in protection from particle degeneracy, just as the flow-based and flow-map \gls*{PF} methods do, these approaches do not provide the flexibility to account for modified particle dynamics between observations which would be reflected in varying importance weights; this is an essential characteristic of the \gls*{nPF} and provides greater ability to adapt to problems with nonlinear dynamics or ``rare" events. 

When $N \ll m$, the distribution $p^N$ is unlikely to have state representations in the tails of $p$, and therefore methods that generate the prior distribution by advecting each particle under the dynamics given by \eqref{equation: basic SDE filtering setup} are likely to result in a prior that does not have support containing the true state in the case of a ``rare" event. 
If instead the particles are allowed to adjust their paths to reflect the most likely path between observations; for instance, by introducing a control $u_t$ to the dynamics
\EquationAligned{
\label{equation: controlled SDE}
dX_t 
= f(X_t) dt + u_t dt + \sigma(X_t) dW_t, 
}
then the posterior distribution at one time can be effectively steered to a representative prior distribution even in the presence of these ``rare" events. 

The approach of \eqref{equation: controlled SDE} is the approach for the \gls*{nPF} \cite{Lingala:2014bu,Yeong:2020} that this work builds upon and will be recalled in \S \ref{section: standard and nudged particle filter}. 
The \gls*{nPF} was originally proposed for mitigating weight degeneracy of the standard \gls*{PF} in nonlinear chaotic and sparse filtering problems, but also has the required formulation to address the ``rare" event problem. 
The central drawback of the \gls*{nPF} is that the exact control solution, which is continuous in time, is often approximated by \gls*{MC} approaches that converge slowly, and hence it is computationally burdensome. 
The aim of this manuscript is to test the efficacy of using a hybrid \gls*{Var-nPF} approach, whereby the variational method is used to generate a pseudo-observation path that provides state-space guidance of the \gls*{nPF} particles. 

As will become clear in the explanations contained in \S \ref{section: variational nudged particle filter}, this allows for a shorter time horizon optimal control problem to be solved for the \gls*{nPF} and hence provides an avenue to reduce its computational cost, while still maintaining the ability to approximate the filtering distribution at all times and handle the case of ``rare" events. 

The paper now proceeds as follows.
In \S \ref{section: standard and nudged particle filter} we recall the basics of the standard bootstrap \gls*{PF} and provide background on the \gls*{nPF}. 
We provide relevant citations to the literature in this section for those methods and set notation. 
In \S \ref{section: variational nudged particle filter} we detail the variational approach used in this work and the new method that combines it with the \gls*{nPF}. 
In \S \ref{section: numerical experiments} we provide the model setup for a stochastic \gls*{L63} system, shown in Fig.~\ref{fig:l63_ic_locations}, that will serve as a surrogate to mimick ``rare" events in geophysical problems.
Simulation results are also given and discussed in this section. 
Final remarks and future efforts are given in \S \ref{section: conclusions}.

\section{Standard and Nudged Particle Filter}
\label{section: standard and nudged particle filter}

\subsection{Standard Particle Filter}
\label{subsection: standard particle filter}

The standard (bootstrap) \gls*{PF} \cite{Gordon:1993ss}, with resampling proceeds in three distinct steps (prior, Bayes' update, and resampling) that repeat for each observation cycle. 
We assume that we start with an $N$-particle approximation of the posterior distribution at time $t_{k - 1}$ with normalized weights $\sum_{i = 1}^N w^i = 1$, as expressed in \eqref{equation: ensemble approximation of a distribution}.
The standard \gls*{PF} with resampling then proceeds recursively by generating the prior at time $t_k$ by advecting each particle $X^i_{k - 1}$ under the dynamics given in \eqref{equation: basic SDE filtering setup} to time $t_k$
\begin{equation}
p_k^N(dx | \mF^Y_{k - 1})
= \sum_{i = 1}^N w^i_{k-1} \delta_{X^i_k}(dx).
\end{equation}
Then, given the observation $Y_k$, unnormalized weights are calculated according to Bayes' formula
\begin{align}
\label{equation: unnormalized weights, standard particle filter}
\wt{w}^i_k 
= w^i_{k - 1} p_k(Y_k | X^i_k),
\end{align}
where $p_k(Y_k | X^i_k)$ is the likelihood function associated with the measurement process given in \eqref{equation: basic SDE filtering setup}. After normalizing these weights $w^i_k = \wt{w}^i_k / \sum_{i = 1}^N \wt{w}^i_k,$
the posterior distribution at time $t_k$ is as 
\begin{equation}
p^N_k (dx | \mF^Y_k)
\equiv
\sum_{i = 1}^N w^i_k \delta_{X^i_k} (dx).
\end{equation}
Lastly, if the particle weights are degenerate (i.e., the variance of the weights is large), then the particles are independently and identically resampled from the particle set $(X^i_k)$ with probability according to the normalized weights $(w^i_k)$.
This process results in particles that may have the same support locations, but now with equal weights (i.e., $w^i_k = 1 / N$).


\begin{figure*}[t]
\begin{center}
\includegraphics[width=140mm]{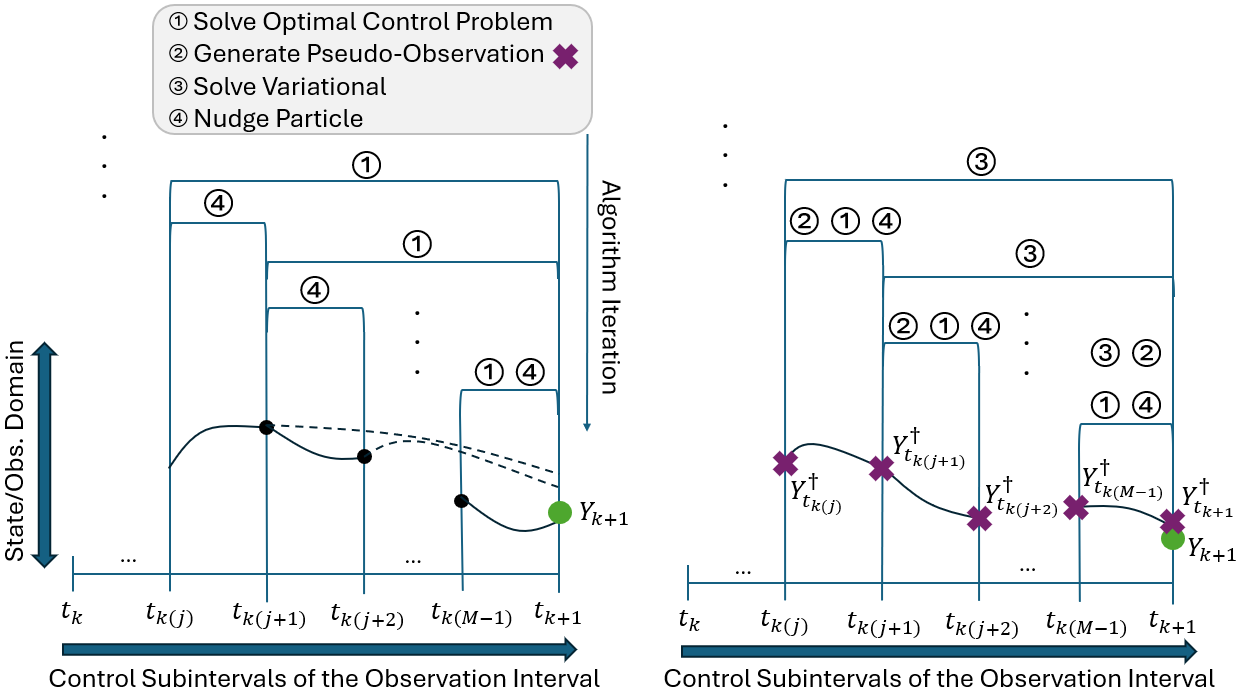}
\end{center}
\caption{Schematic evolution of a single particle within one observation interval for the \gls*{nPF} (left) and the \gls*{Var-nPF} (right).}
\label{figure:Caricature_figure}
\end{figure*}

To determine whether the particle approximation is degenerate in this work, we use an approximation of the \gls*{ESS} \cite{Bergman:1999,Liu:1998jasa}. 
With the vector of normalized weights denoted as $w$, the approximation is given by 
\begin{align*}
N_{\textrm{eff}}
\equiv 1 / \langle w, w \rangle.    
\end{align*}
We use the threshold of $N_{\textrm{eff}} < N / 2$ to trigger resampling with the universal (or systematic) resampling approach \cite{Kitagawa:1996jcgs}. 

\subsection{Nudged Particle Filter}
\label{subsection: nudged particle filter}

The key difference between the \gls*{nPF} and a standard \gls*{PF} algorithm is that the particles are advected under the dynamics given by \eqref{equation: controlled SDE} as opposed to \eqref{equation: basic SDE filtering setup}. 
The forcing $u_t$ is a nonlinear control law that is obtained independently for each particle $X^i$ by solving a finite-time horizon \gls*{OCP} that is dependent on the new observation. 
The result of applying the control is that particles are \emph{nudged} toward the area of high likelihood for the new observation. 
The setup and main result of this \gls*{OCP} are given in the following theorem statement. 

\paragraph*{Notation}
The superscript notation $X^{k, x}$ for \emph{any} particle indicates an initial condition at $x \in \R{m}$ at time $t_k$. 
$C([t_k, t_{k +1}])$ is the space of continuous functions and a superscript asterisk (e.g., $\sigma^*$) indicates the transpose of a matrix. 

{
\begin{theorem}[\textcolor{\statementcolor}{Nudged Particle Filter (nPF)} \cite{Lingala:2014bu, Yeong:2020}]
\label{theorem: nudged particle filter}
Given a sample $X^{k, x}_{t_k} \sim p^N_k(dx | \mF^Y_k)$ and observation $Y_{k + 1}$, the solution of the \gls*{OCP}
\begin{multline}
\label{eq:opt_ctrl_problem}
\min_{u \in C([t_k, t_{k +1}])} \left\{ 
J(u; k, x, Y_{k + 1}) \right.  \\
\left. \equiv \E{ \int_{t_k}^{t_{k + 1}} \frac{1}{2} \langle u_s, R^{-1}_s u_s \rangle ds + g(X^{k, x}_{k + 1}, Y_{k + 1}) } 
\right\},
\end{multline}
with dynamic constraint for $X^{k, x}_t$, $t \in [t_k, t_{k + 1})$ given by \eqref{equation: controlled SDE}, 
$g$ the log-likelihood of the observation, $R_s$ the diffusion matrix of the stochastic forcing in \eqref{equation: controlled SDE}, 
is (under reasonable assumptions) a feedback control law
\begin{align}
\label{equation: feedback control solution of log-transformed HJB}
u(t, x) = \frac{1}{\Phi(t, x)} R_t \nabla_x \Phi(t, x),
\end{align}
with $\Phi$ and $\nabla_x \Phi$ solutions to linear parabolic PDEs that can be written in terms of Feynman-Kac relations
\begin{align}
\label{equation: Feynman-Kac application for log-transformation solution}
\Phi(t, x) 
= \E{ \exp( - g(\eta^{k, x}_{k + 1}, Y_{k + 1})) },  
\end{align}
and
\begin{multline}
\label{equation: Feynman-Kac application for log-transformation solution of derivative}
\nabla_x \Phi(t, x) 
= -\mathbb{E} \left[ \exp( - g(\eta^{k, x}_{k + 1}, Y_{k + 1})) \right. \\
\left. \exp \left( \int_{t_k}^{t_{k + 1}} \nabla_x f(\eta^{k, x}_s) ds \right) \nabla_x g(\eta^{k, x}_{k + 1}, Y_{k +1}) \right],
\end{multline}
with $\eta^{k, x}_{k + 1}$ being realizations of \eqref{equation: basic SDE filtering setup}.
\end{theorem}
}

An important result of the control solution given in \eqref{equation: feedback control solution of log-transformed HJB} is that particles attempt to stay faithful to the modeled dynamics of \eqref{equation: basic SDE filtering setup}.
Any deviation is accounted for by a change in their unnormalized weights. 

{
\begin{corollary}[\textcolor{\statementcolor}{nPF Radon-Nikodym Derivative} \cite{Yeong:2020}]
\label{corollary: nPF Radon-Nikodym derivative}
The unnormalized weights of the sample $X^{k, x}_t$ under the control law \eqref{equation: feedback control solution of log-transformed HJB} given by Theorem \ref{theorem: nudged particle filter} change according to the continuous-time Radon-Nikodym derivative
\begin{multline}
\label{equation: radon-nikodym derivative correction}
\restrict{\frac{d\mu^i}{d\widehat{\mu}^i}}{t \in [t_k, t_{k + 1}]}
= \exp \left( -\int_{t_k}^t \langle v(s, X^i_s) , dW_s \rangle \right. \\
\left. - \frac{1}{2} \int_{t_k}^t \langle v(s, X^i_s) , v(s, X^i_s) \rangle ds \right),
\end{multline}
where $v(s, X^i_s) = \sigma^* \nabla_x \left(\log \Phi(s, X^i_s) \right)$ when $R \equiv \sigma \sigma^*$.
Therefore, in contrast to \eqref{equation: unnormalized weights, standard particle filter}, the unnormalized weights of the \gls*{nPF} are 
\begin{align}
\label{equation: unnormalized weights, nudged particle filter}
\wt{w}^i_{k + 1} = \restrict{\frac{d \mu^i}{d \widehat{\mu}^i}}{t_{k + 1}} w^i_{k} p_k(Y_{k + 1} | X^i_{k + 1}).
\end{align}
\end{corollary}
}
For additional details on the \gls*{nPF}, more advanced versions, and its application to higher-dimensional mathematical models than given in this manuscript, we direct the reader to some of the recent literature \cite{Beeson:2020nd, Beeson:2025.fusion.147.irnpf, Zhou:2024.fusion}. 

\subsubsection{Adaptive Nudging Calculation}
\label{subsubsection: adaptive nudging calculation}

Numerical implementation of the \gls*{nPF} requires an approximation of \eqref{equation: Feynman-Kac application for log-transformation solution} and \eqref{equation: Feynman-Kac application for log-transformation solution of derivative}, which is most directly accomplished via \gls*{MC}. 
To ensure that our approximation of the expectations and hence the control solution has converged to an acceptable tolerance level, an adaptive scheme is used whereby we calculate the control for an increasing number of realizations $(\eta^{k, x}_{k + 1})$.
This is implemented in batch form to improve efficiency and reuse previously computed realizations.
In particular, let $K \in \N{}$ be a batch size and $j \in \N{}$ the $j$-th iteration of the algorithm to compute the control approximation $u^{j K}_t$, with $jK$ total realizations. 
Additionally, let 
\begin{equation}
    \hat{u}^{j K}_t \equiv u^{j K}_t / |\E{f(X_t)}|
\end{equation}
be the normalized control magnitude. 
Repeating this approximation for the $(j + 1)$-th iteration, we define the normalized control variation to be
\begin{equation}
    \delta \hat{u}^{j + 1}_t \equiv | \hat{u}^{(j + 1) K}_t - \hat{u}^{j K}_t|.
\end{equation}
Then for a user specified tolerance $\epsilon_u > 0$, if $\delta \hat{u}^{j + 1} \leq \epsilon_u$, the approximate control solution $\hat{u}^{(j + 1) K}_t$ is said to have converged, and otherwise we proceed to the $(j + 2)$-th iteration of the algorithm. 

\subsubsection{Additional Remarks}
\label{subsubsection: additional remarks}

We finish this section by making clear the distinction between the \gls*{nPF} and flow-based \gls*{PF} algorithms mentioned in \S \ref{section: introduction}, and in particular those derived also from a control perspective such as the \gls*{FPF} \cite{Yang:2013.ieee.tac.58.10}. 
The main distinction is that the \gls*{FPF} applies a control law in the pseudo-observation interval for the Bayes' update, whereas the \gls*{nPF} applies a control to the particles between observations. 
Since the control in the \gls*{nPF} occurs between observations, the weights must be changed accordingly. 

Another approach that is closely related to the \gls*{nPF}, but motivated by large deviation theory and derived from a variational method approach, has been applied to a model for the rare event of the Kuroshio current \cite{Vanden-Eijnden:2013.mwr.141.6}.

\section{Variational Nudged Particle Filter}
\label{section: variational nudged particle filter}

It is not practical to solve for a continuous-time control of the \gls*{nPF} at every time $t \in [t_k, t_{k + 1})$ between observations. 
Therefore, the standard implementation is to partition the interval between observations into $M \in \N{}$ equal-length control subintervals with a collection of endpoints defining them given by the indexed set $(t_{k(j)})_{j = 0}^M$. 
Then a control solution is found for the time $t_{k(j)}$ at the beginning of each subinterval, by solving the nonlinear \gls*{OCP} in (\ref{eq:opt_ctrl_problem}) from $t_{k(j)}$ to $t_{k+1}$, and this control is then held constant in application for the remainder of the subinterval $[t_{k(j)}, t_{k(j + 1)})$. 
As the expectations in \eqref{equation: Feynman-Kac application for log-transformation solution} and \eqref{equation: Feynman-Kac application for log-transformation solution of derivative} are over the state $\eta^{k, x}_{k + 1}$ at the final time, the earlier control subintervals are computationally more expensive (i.e., longer numerical integration times) than the later ones; this is shown in Fig.~\ref{figure:Caricature_figure} (left). 
Additionally, the earlier subintervals are more sensitive to the nonlinear dynamics and noise perturbations, which requires larger batch sizes in the adaptive nudging calculation described in \S \ref{subsubsection: adaptive nudging calculation}. 

The central contribution of this paper is to introduce a pseudo-observation path that enables a fixed computational cost for the nudging control in every subinterval. 
This pseudo-observational path also appears to help moderate the weight variance due to the Radon-Nikodym derivative for the \gls*{nPF} as given in Corollary \ref{corollary: nPF Radon-Nikodym derivative}. 

For clarity of discussion, let us fix a single observation interval $[t_k, t_{k + 1})$ with observation $Y_{k + 1}$ and particles $(X^i_k)$ (refer to Fig.~\ref{figure:Caricature_figure} (right)).  
We then aim to find the most likely path of the observation; that is a path $(Y^\dagger_t)_{t \in [t_k, t_{k + 1}]}$ such that $Y^\dagger_{t_{k + 1}} \approx Y_{k + 1}$. 
This likely observation path can then be substituted into \eqref{equation: Feynman-Kac application for log-transformation solution} and \eqref{equation: Feynman-Kac application for log-transformation solution of derivative} so that control solutions at each subinterval can be calculated with realizations ending at the end of the subinterval, $t_{k(j+1)}$, instead of the full observation interval, $t_{k+1}$. 
To find such a likely observation path, we instead solve for a state-space path $(X^\dagger_t)$ that is the most likely path from a given initial condition to a state that generates the observation $Y_{k + 1}$. 
In particular, this state-space path is found by solving a variational problem. 
Specifically, what is often referred to as the \gls*{sc4DVar} approach in the geophysical data assimilation community \cite{Evensen:2022.springer.international, Talagrand:1987.qjrms.113.478, Courtier:1987.qjrms.113.478}. 
The optimization problem to be solved is
\EquationAligned{
\label{equation: variational cost function}
\min_{x \in \R{m}} \left\{ 
J(x; k, Y_{k + 1}) 
\equiv 
\frac{1}{2} \langle \Delta x, \left( \Sigma^N_k \right)^{-1} \Delta x \rangle 
\qquad \quad \right. \\ 
\left. 
+ \frac{1}{2} \langle \Delta y , \Sigma_y^{-1} \Delta y \rangle
\right\},
}
with $\mu^N_{k}$ and $\Sigma^N_{k}$ an empirical mean and covariance respectively (e.g., for the \gls*{nPF} solution) at time $t_{k}$, $\Delta x \equiv x - \mu^N_{k}$, 
$\Delta y \equiv Y_{k + 1} - h(X^{{k}, x}_{k + 1})$ is the innovation, where $X^{{k}, x}_{k + 1}$ is the solution of the deterministic dynamics (i.e., $\sigma = 0$ in \eqref{equation: basic SDE filtering setup}). 
If $(X^\dagger_t)$ is now the most likely path integrated under the deterministic dynamics, starting from an optimal solution to \eqref{equation: variational cost function}, then the pseudo-observation path is simply $Y^\dagger_t \equiv h(X^\dagger_t)$. 

In the case where the measurement operator is simply the identity mapping (i.e., $h$ is the identity matrix $\op{Id.} \in \R{m \times m}$), then the observation and state-space are identical, and we can use $X^\dagger_t$ as the pseudo-observation path itself.  
We note that it is natural to also extend \eqref{equation: variational cost function} to the case of multiple observation at different times, but in this work we only consider assimilation cycles with one observation time. 

\subsection{The Variational Nudged Particle Filter Algorithm}
\label{subsection: the variational nudged particle filter algorithm}

The combination of the variational approach with the \gls*{nPF} now proceeds by first computing a solution to the variational problem of \eqref{equation: variational cost function} for a given control subinterval described in \S \ref{section: variational nudged particle filter}.
Advecting the a deterministic state starting from this optimal initial condition to the end of the control subinterval generates a target pseudo-observation $Y^\dagger_{t_{k(j + 1)}}$ that is then used as a conditional parameter (pseudo-observation) to the \gls*{nPF} \gls*{OCP} in Theorem \ref{theorem: nudged particle filter}. 
These pseudo-observations are shown in Fig.~\ref{figure:Caricature_figure} (right). 

As discussed at the beginning of this section, the nudged control problem is now always of fixed time duration, being equal to the subinterval length $t_{k(j)} - t_{k(j+1)}$. 
We repeat this procedure for each control subinterval, including the final one, where we still use the produced pseudo-observation $Y^\dagger_{t_{k + 1}}$ instead of the true given one $Y_{k + 1}$.
This new algorithm is labeled \gls*{Var-nPF} throughout the remainder of the paper.  

For the numerical experiments given in \S \ref{section: numerical experiments}, the optimization of the variational problem given by \eqref{equation: variational cost function} is solved with a bounded limited Broyden-Fletcher-Goldfarb-Shannon (L-BFGS-B) \cite{Scipy:Optimize:2025:07:06}. 
In the case that $\Sigma^N_k$ is poorly conditioned, we add a small regularization term $\epsilon \op{Id.}$, with $0 < \epsilon \ll 1$. 

\section{Numerical Experiments}
\label{section: numerical experiments}

In this section we first explain the setup of a stochastic version of the \gls*{L63} model \cite{Lorenz63} to be used in experiments, then pertinent simulation parameters, and lastly the findings and comparisons of individual and \gls*{MC} experiments.  

\subsection{The Stochastic Lorenz 1963 Model}
\label{subsection: the stochastic lorenz 1963 model}

The deterministic definition of the \gls*{L63} model \cite{Lorenz63} has a drift vector field $f$ for \eqref{equation: basic SDE filtering setup} given by 
\begin{align}
f: (x, y, z) \mapsto (\alpha (y-x), \gamma x - y - xz, xy - \beta z), 
\end{align}
where standard parameters are $\alpha = 10, \gamma = 28, \beta = 8/3$, and are used in this work. 
With these parameters the \gls*{L63} model is chaotic and solutions converge onto an attractor as shown in Fig.~\ref{fig:l63_ic_locations}.
The error doubling time, an indicator of how chaotic a system is, is approximately 0.76 \gls*{TU}, and the average time that a solution switches from one lobe of the attractor to the other is approximately 1.75 \gls*{TU}. 

The \gls*{nPF} requires a stochastic dynamical system for its application. 
We therefore modify the \gls*{L63} model by adding the stochastic forcing shown in \eqref{equation: basic SDE filtering setup}. 
In particular, we consider the additive noise case, where the dispersion coefficient $\sigma$ is constant and defined so that the diffusion coefficient for the stochastic forcing is
\begin{align}
\sigma \sigma^* 
\equiv
\begin{bmatrix}
2 & 1 & 0.5\\
1 & 2 & 1\\
0.5 & 1 & 2
\end{bmatrix}.
\end{align}


\subsection{Simulation Parameters}
\label{subsection: simulation parameters}

All simulations will use $N = 10$ particles, observation step-size $\Delta t_{\text{obs}} = 0.5$~\gls*{TU}, and final time $t_f = 3.5$~\gls*{TU}. 
Numerical intergration is with a Wiener-RK4-Maruyama scheme, with integrator step-size $\Delta t = 0.01$~\gls*{TU}. 
Unless stated otherwise, the initial condition is chosen as $x = ( 1.508870, -1.531271, 25.46091 )$, which is identified in Fig. \ref{fig:l63_ic_locations} by the \textcolor{red}{red star}. 

The initial particle ensembles for any \gls*{PF} method is via sampling from a Gaussian $\mathcal N(\mu, 2 \op{Id.})$, where $\mu$ represents one of the initial conditions shown in Table \ref{tab:mc_ic_sweep_ess_rmse} and labeled on the attractor in Fig. \ref{fig:l63_ic_locations}.
These representative initial conditions were generated from a long deterministic trajectory of the attractor with the aim to provide low-discrepancy samples of the attractor itself. 
These initial conditions therefore provide dynamically distinct states and promote \gls*{MC} results that provide a fairer benchmarking of the various filtering methods. 

The observation operator is $h = \op{Id.}$ with noise covariance $\Sigma_y = 2 \op{Id.}$, so that the standard deviation of the observation noise 
is less than $5\%$ of the characteristic range of the \gls*{L63} state variables. 
This ensures that the observations are informative without dominating the model dynamics. 

Each observation interval is partitioned into $M = 5$ control subintervals. 
In the solution of the \gls*{OCP} for each subinterval,  
we use $K = 2$ trajectory realizations per batch together with the adaptive nudging strategy described in \S \ref{subsubsection: adaptive nudging calculation}. 
A tolerance of $\epsilon_u = 0.1$ is used for the adaptive nudging calculation. 
A maximum of 50 batches (i.e., 100 total realizations) is allowed in the calculation, though none of the experiments required this many to converge to our tolerance of $\epsilon_u$. 


\subsection{Simulation Results}
\label{subsection: simulation results}


We first consider a fixed simulation scenario to illustrate the prototypical behavior of the \gls*{PF}, \gls*{nPF}, and \gls*{Var-nPF} methods. 
Then having established and visualized the behavior of these methods, we perform a \gls*{MC} analysis.

\paragraph*{Prototypical Behavior} 

\begin{figure}[t]
\centering
\includegraphics[width=\columnwidth]{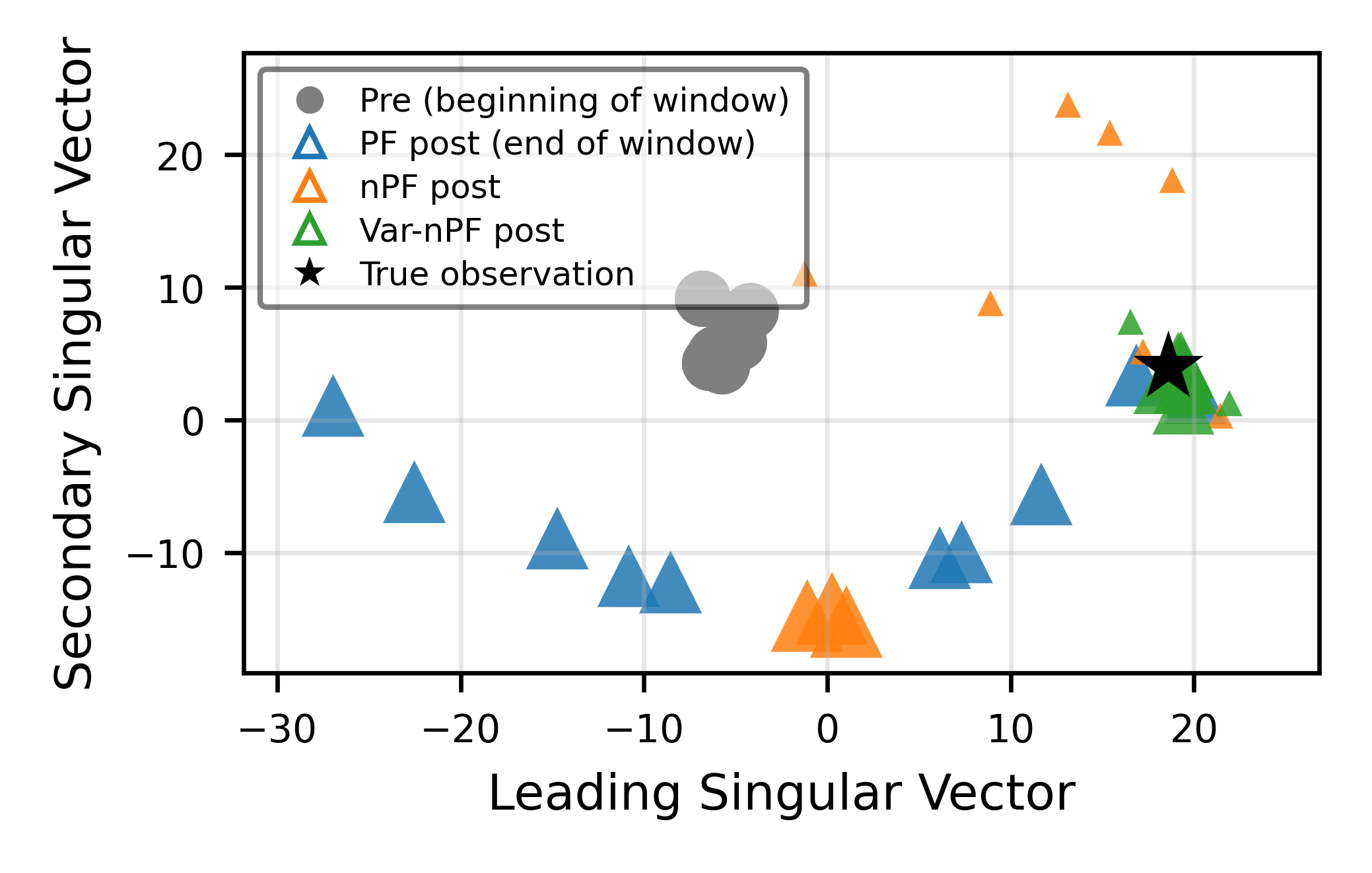}
\caption{
Initial particle samples for all methods shown with \textcolor{gray}{gray circle markers}. 
Prior states of each method (triangles) at the first observation $t = 0.5$ scaled to the normalized weight. 
}
\label{fig:pre_post_all_variants_PCA}
\end{figure}

To better understand and visualize particle motion within a single observation interval across the three \gls*{PF} variants presented in this work, we consider the case where all filters use the same samples generated from a Gaussian distribution with mean given by the \textcolor{red}{red star} in Fig. \ref{fig:l63_ic_locations}.
These samples are shown by \textcolor{gray}{gray circle markers} in Fig. \ref{fig:pre_post_all_variants_PCA}, where they have been projected onto the dominate subspace spanned by the prior states of all the particles (i.e., all the methods) at the first observation time. 
This subspace is found by a singular value decomposition of a matrix with all the particle states as columns. 

\begin{figure}[t]
\centering

\begin{subfigure}[t]{\linewidth}
  \centering
  \includegraphics[width=\linewidth]{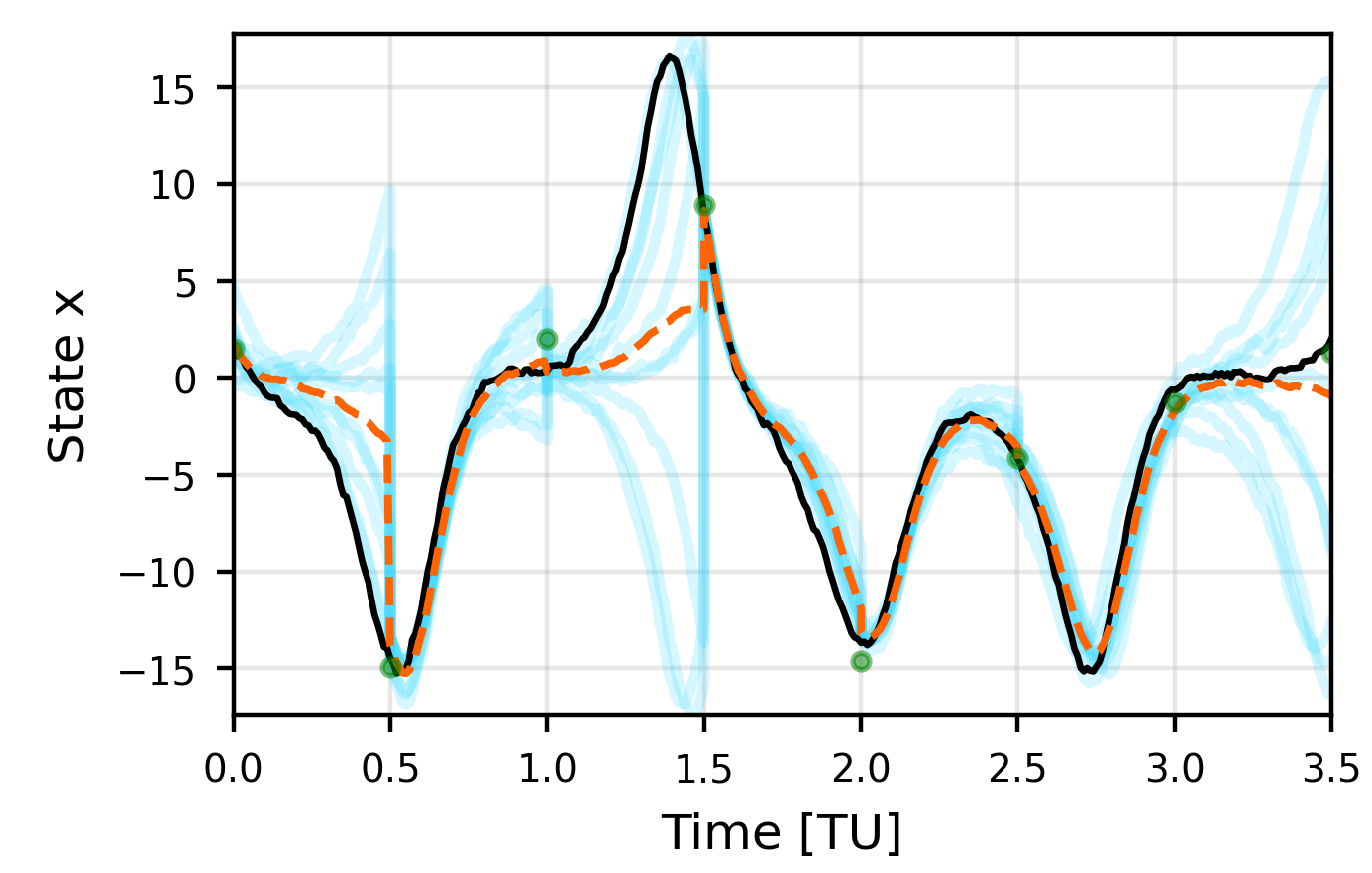}
  \caption{PF}
\end{subfigure}

\vspace{0.6em}

\begin{subfigure}[t]{\linewidth}
  \centering
  \includegraphics[width=\linewidth]{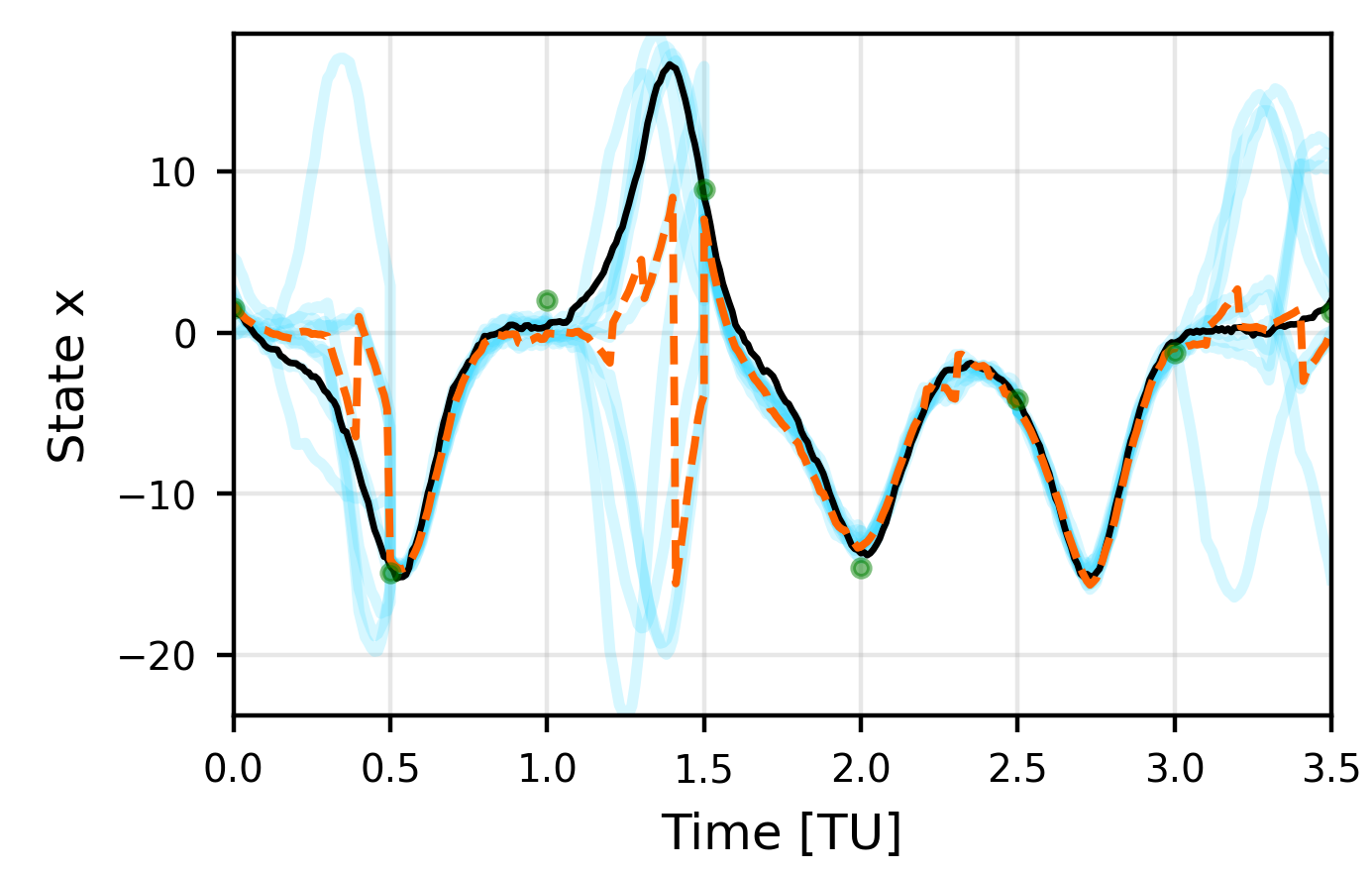}
  \caption{nPF}
\end{subfigure}

\vspace{0.6em}

\begin{subfigure}[t]{\linewidth}
  \centering
  \includegraphics[width=\linewidth]{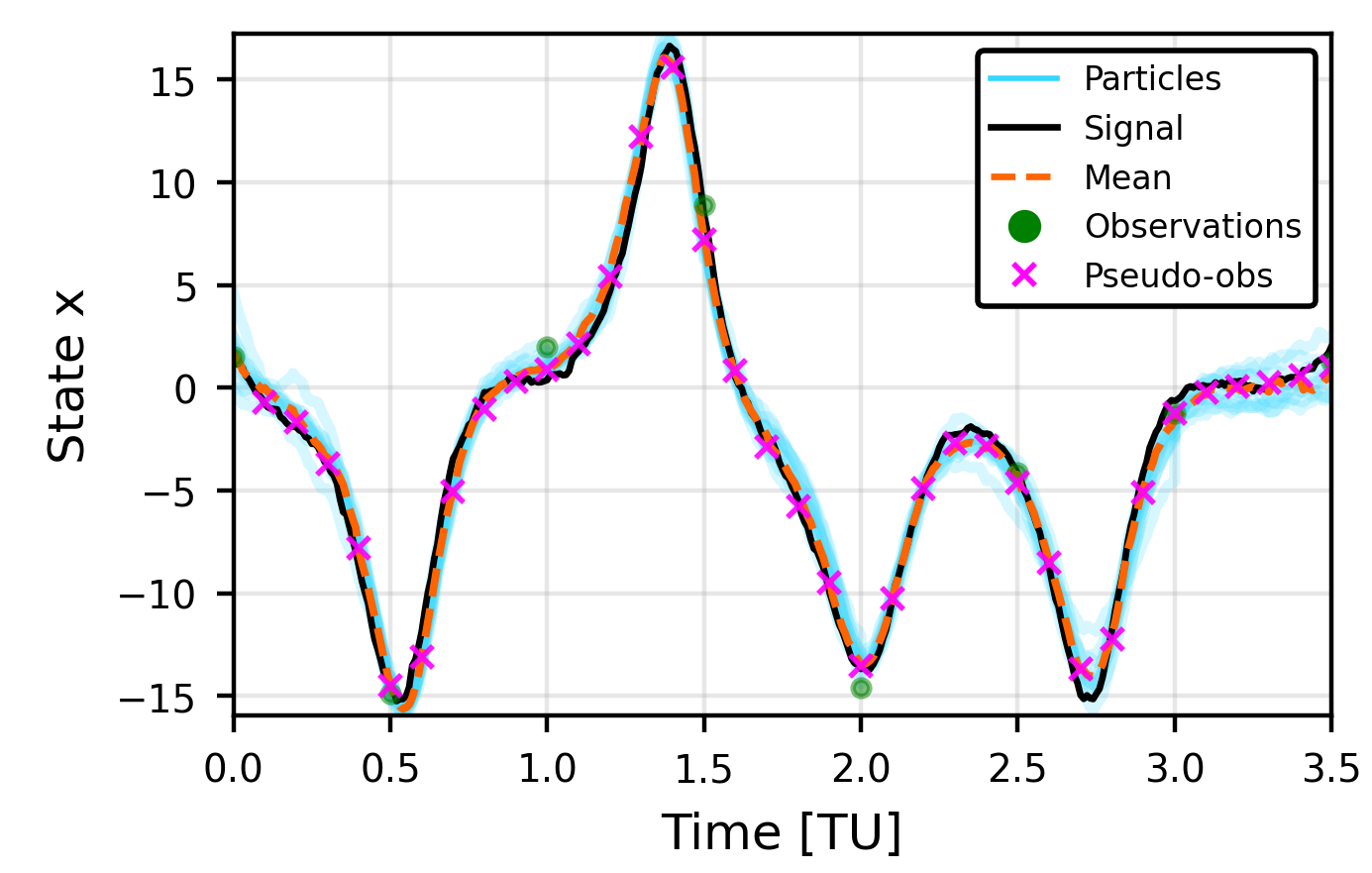}
  \caption{Var-nPF}
\end{subfigure}

\caption{Time evolution of the first state component (i.e., $x$).}
\label{fig:State_X_versus_time}
\end{figure}

All three methods perform filtering with the same true hidden signal (\textbf{black} line) and observations (\textcolor{ForestGreen}{green} markers) as shown in Fig. \ref{fig:State_X_versus_time}. 
Fig. \ref{fig:pre_post_all_variants_PCA} shows the prior particle locations with triangular markers, and the size of the markers is a reflection of each particles normalized weight. 
In particular, the \gls*{PF} does not change the weights of the particles, whereas the \gls*{nPF} and \gls*{Var-nPF} weights do change as the particles attempt to nudge closer to the observation. 
The \gls*{PF} \gls*{nESS} changes from 1 for the prior to 0.15 after the Bayes' update, where as the \gls*{nPF} goes from 0.3 to 0.1, and the \gls*{Var-nPF} from 0.56 to 0.41. 



While the \gls*{PF} spreads significantly after advection, the \gls*{nPF}, and especially those of the \gls*{Var-nPF}, move toward the observation. 
This observation interval corresponds to a ``rare" event (i.e., a transition of the lobes), as shown in Fig. \ref{fig:State_X_versus_time}. 
It is for this reason that the \gls*{PF}, which has no control, ends up with too much variance at the time of the observation. 
It is promising that all of the \gls*{Var-nPF} particles arrive in the area of high likelihood for the observation, and most with reasonably good weights. 
We see this behavior also holds true for other switching times between lobes in Fig. \ref{fig:State_X_versus_time}.

The \gls*{nPF} particles that do not arrive near the observation is due to our implementation that prevents nudging to occur if it would reduce an individual particles unnormalized weight too much. 
We refer to such cases in this paper as a ``zero control rollback" (i.e., we calculate the proposed control, but ultimately do not apply it\textendash see Table \ref{tab:control_regularization}). 

\begin{figure}[t]
\centering

\begin{subfigure}[t]{\linewidth}
  \centering
  \includegraphics[width=\linewidth]{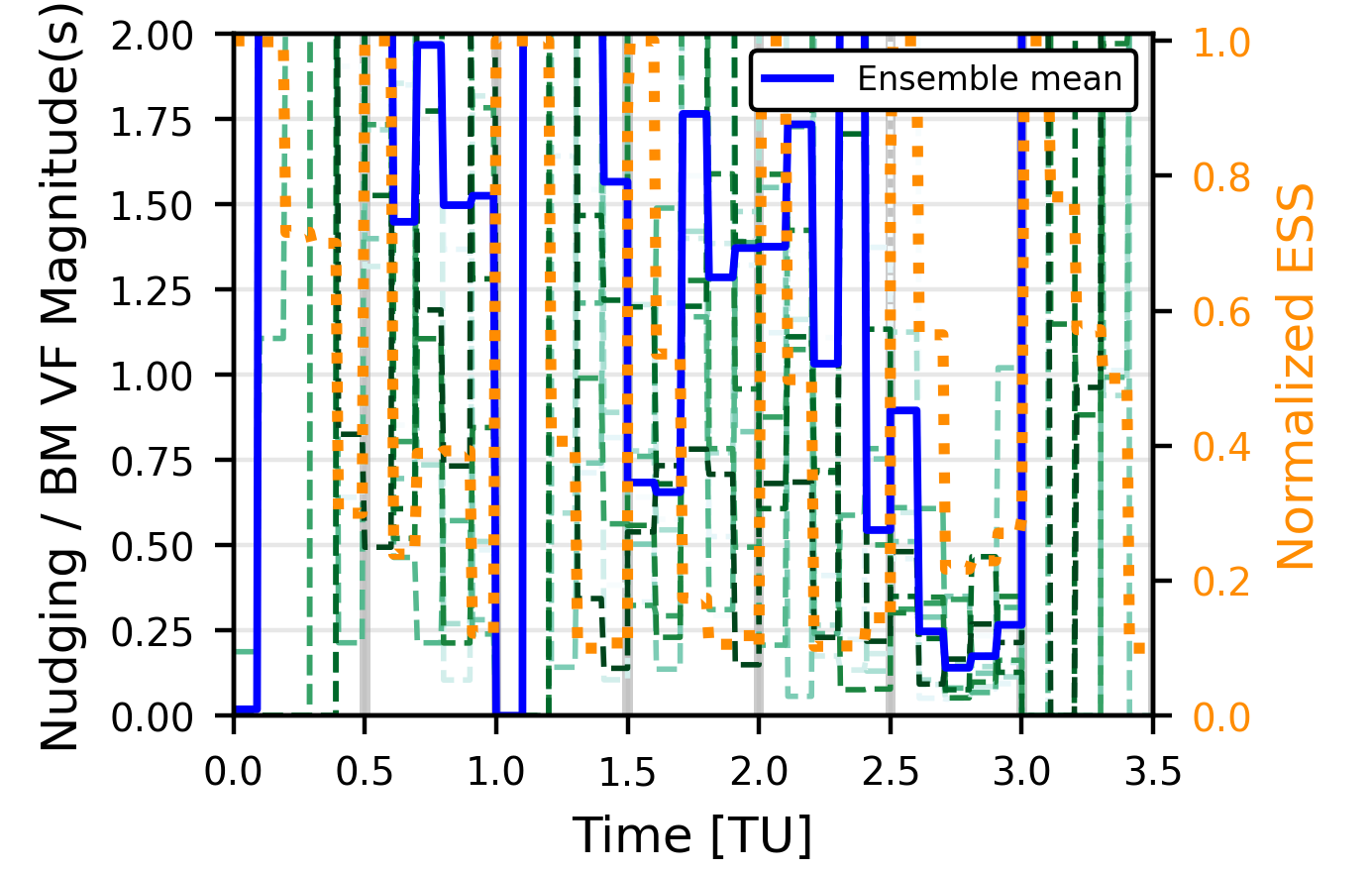}
  \caption{nPF}
\end{subfigure}

\vspace{0.6em}

\begin{subfigure}[t]{\linewidth}
  \centering
  \includegraphics[width=\linewidth]{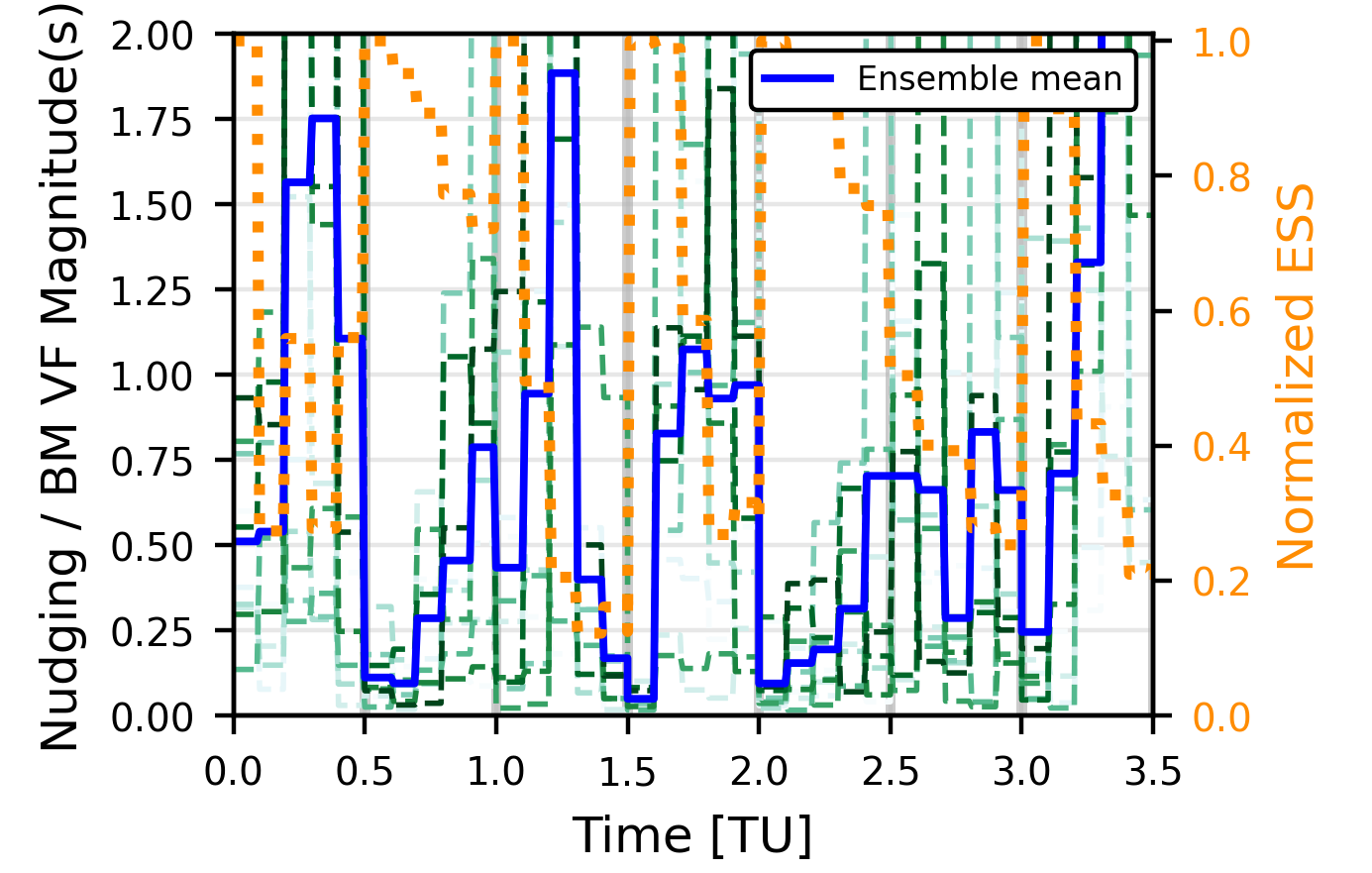}
  \caption{Var-nPF}
\end{subfigure}

\caption{
Nudging magnitude relative to \gls*{BM} forcing (\textcolor{ForestGreen}{green}), ensemble mean for this relative control (\textcolor{blue}{blue}), and the \gls*{nESS}. 
}
\label{fig:nudging_vs_BM}
\end{figure}

It is interesting to compare the nudging control magnitude relative to the \gls*{BM} \gls*{VF} as shown in Fig. \ref{fig:nudging_vs_BM} for this experiment. 
This relative nudging to \gls*{BM} \gls*{VF} is shown in \textcolor{ForestGreen}{green} for each particle and the ensemble mean in \textcolor{blue}{blue}. 
The \gls*{nESS} is shown with the second y-axis in \textcolor{orange}{orange}. 

Of particular note is the aggressive control of the \gls*{nPF}, whereas our new method is able to relax the proposed relative control and hence deliver more particles to the observation location and with better \gls*{nESS}. 
This is most easily observed from the ensemble mean of the nudging relative to the \gls*{BM} \gls*{VF} taking larger values (above the upper bound of the plot) for the \gls*{nPF}, but only at the end of the simulation for the \gls*{Var-nPF}. 
These occurrences correspond to larger dispersion in the particles at those times, which can be seen in Fig. \ref{fig:State_X_versus_time}.
It is important to note that the nudging to \gls*{BM} \gls*{VF} ratio should remain below a value of one on average if we are to believe that the stochastic model is properly defined (i.e., if the stochastic forcing represents model uncertainty, then it is parameterized properly to reflect reality and our filtering method is not trying to overpower the model defined dynamics). 
We indeed see this to happen more often in the new \gls*{Var-nPF}, which is a promising indicator.

The computational overhead introduced by the variational component was also analyzed for this experiment and \gls*{MC} runs, which showed similar trends. 
In particular, solving the variational problem accounts for approximately 40.5\% of the total \gls*{Var-nPF} runtime. 
The remaining run time was due to the \gls*{nPF} control calculations. 
Although the variational component constitutes a significant portion of the computation, it does not dominate the overall runtime and reduces the computational cost compared to the \gls*{nPF} for this experiment (see Table \ref{tab:control_regularization}). 
As long as a decent optimization solution is provided by the variational part of the new method, it has the potential to scale much better for higher dimensional and longer observation interval problems. 
The average \gls*{RMSE} for each method shown in Table \ref{tab:control_regularization}, is the average over the state and time with respect to the true signal. 

\begin{table}[t]
\caption{
Comparison between \gls*{PF}, \gls*{nPF}, and \gls*{Var-nPF} for one experiment with the \textcolor{red}{red star} initial mean condition.
}
\label{tab:control_regularization}
\centering
\footnotesize
\setlength{\tabcolsep}{4pt}
\renewcommand{\arraystretch}{1.1}
\begin{tabular}{|l|c|c|}
\hline
\textbf{Metric} & \textbf{nPF} & \textbf{\gls*{Var-nPF}} \\
\hline
Mean of Control Magnitude $\|u\|_2$ & 18.14 & 4.55 \\
\hline
Maximum of Control Magnitude $\|u\|_2$ & 220.91 & 41.13 \\
\hline
Fraction of Zero Control Rollback & 0.166 & 0.000 \\
\hline
Average Nudging / \gls*{BM} VF ratio & 2.87 & 0.77 \\
\hline
Maximum Nudging / \gls*{BM} VF ratio & 60.62 & 9.69 \\
\hline
Average \gls*{RMSE} & 4.92 & 1.48 \\
\hline
Runtime (s) & 13.49 & 7.8 \\
\hline
\end{tabular}
\end{table}

\paragraph*{Monte Carlo Simulations} 

\begin{table}[t]
\caption{100 \gls*{MC} runs; \textcolor{red}{ted star} initial mean in Fig. \ref{fig:l63_ic_locations}}
\begin{center}
\begin{tabular}{|c|c|c|c|c|c|}
\hline
\textbf{Metric} & \multicolumn{5}{|c|}{\textbf{Filter Type}} \\
\cline{2-6}
 & \textbf{\textit{PF$_{10}$}} & \textbf{\textit{PF$_{30}$}} & \textbf{\textit{PF$_{40}$}} & \textbf{\textit{nPF}} & \textbf{\textit{\gls*{Var-nPF}}} \\
\hline
Number of Particles & 10 & 30 & 40 & 10 & 10 \\
\hline
Avg. \gls*{RMSE} & 6.35 & 4.47 & 3.66 & 6.39 & 2.91 \\
\hline
Normalized Avg. \gls*{ESS} & 0.27 & 0.26 & 0.27 & 0.17 & 0.26 \\
\hline
Runtime (s) & 18.47 & 30.45 & 37.21 & 283.9 & 132.92 \\
\hline
\end{tabular}
\label{tab:mc_summary_with_pf3}
\end{center}
\end{table}

\begin{table}[t]
\caption{
\gls*{MC} (1000 total runs, 100 for each $\mu$ initial condition),  
reporting Avg. \gls*{nESS}, Avg. \gls*{RMSE}, and runtime. 
}
\label{tab:mc_ic_sweep_ess_rmse}
\centering
\footnotesize
\setlength{\tabcolsep}{4pt}
\renewcommand{\arraystretch}{1.15}
\resizebox{\columnwidth}{!}{%
\begin{tabular}{|c|c|cc|cc|cc|}
\hline
\textbf{\#} &
\textbf{$\mu$ Initial Condition} &
\multicolumn{2}{c|}{\textbf{PF}} &
\multicolumn{2}{c|}{\textbf{nPF}} &
\multicolumn{2}{c|}{\textbf{Var-nPF}} \\
\hline
 &  &
\textbf{\gls*{nESS}} & \textbf{\gls*{RMSE}} &
\textbf{\gls*{nESS}} & \textbf{\gls*{RMSE}} &
\textbf{\gls*{nESS}} & \textbf{\gls*{RMSE}} \\
\hline
*  & $(1.509,\,-1.531,\,25.461)$   & 0.27 & 6.35 & 0.17 & 6.39 & 0.26 & 2.91 \\ \hline
1  & $(-3.622,\,2.487,\,29.784)$   & 0.26 & 6.64 & 0.14 & 8.34 & 0.25 & 4.37 \\ \hline
2  & $(-8.587,-14.288,\,16.895)$   & 0.30 & 6.83 & 0.20 & 6.43 & 0.27 & 4.64 \\ \hline
3  & $(-14.411,-8.058,\,40.440)$   & 0.26 & 6.50 & 0.17 & 5.79 & 0.24 & 4.45 \\ \hline
4  & $(14.418,\,11.236,\,37.915)$  & 0.25 & 7.96 & 0.18 & 5.69 & 0.24 & 6.17 \\ \hline
5  & $(4.133,\,6.815,\,14.316)$    & 0.28 & 8.02 & 0.19 & 7.75 & 0.26 & 5.35 \\ \hline
6  & $(-2.895,-5.123,\,11.843)$    & 0.32 & 5.57 & 0.19 & 5.47 & 0.27 & 3.98 \\ \hline
7  & $(-5.802,-7.589,\,20.507)$    & 0.36 & 4.00 & 0.22 & 3.70 & 0.31 & 2.94 \\ \hline
8  & $(10.347,\,17.701,\,17.250)$  & 0.25 & 8.86 & 0.16 & 7.42 & 0.24 & 5.41 \\ \hline
9  & $(3.072,-0.052,\,26.056)$     & 0.25 & 7.96 & 0.14 & 8.80 & 0.25 & 4.28 \\ \hline
10 & $(1.909,-0.842,\,24.846)$     & 0.22 & 10.31 & 0.17 & 6.76 & 0.26 & 2.98 \\ \hline
\textbf{Avg.} & --- 
& \textbf{0.28} & \textbf{7.27}
& \textbf{0.18} & \textbf{6.62}
& \textbf{0.26} & \textbf{4.46} \\ 
\hline
\textbf{Avg.} &
\textbf{Computational Runtime} &
\multicolumn{2}{c|}{\textbf{23.39}} &
\multicolumn{2}{c|}{\textbf{202.08}} &
\multicolumn{2}{c|}{\textbf{129.87}} \\
\hline
\end{tabular}%
}
\end{table}

To validate that the trends and observations seen in the single experiment described in the last section hold in general, we performed a \gls*{MC} analysis of 100 simulations with the \textcolor{red}{red} star initial condition (see Fig. \ref{fig:l63_ic_locations}) as the mean of a Gaussian to be sampled.
These \gls*{MC} results are given in Table \ref{tab:mc_summary_with_pf3} and it shows that the \gls*{nPF} improves the average RMSE slightly over the \gls*{PF}, but with degradation in the \gls*{nESS}. 
The limiting factor in the performance of the \gls*{nPF} to achieve better \gls*{RMSE} is due to the fact that we only use $M = 5$ control subintervals, which therefore results in the non-insignificant fraction of zero control rollbacks detailed in Table \ref{tab:control_regularization}. 
Using more control subintervals would drastically increase the runtime of the \gls*{nPF}, which was already deemed high for comparisons to the \gls*{PF} for this low-dimensional model. 

There is noticable improvement in the the \gls*{RMSE} for the new \gls*{Var-nPF} without a degradation in the \gls*{nESS}. 
Table \ref{tab:control_regularization} also shows \gls*{MC} results for the \gls*{PF} with an increasing number of particles, so that a better comparison of the performance of the methods on an estimation accuracy and computational runtime level can be made. 
Although the \gls*{Var-nPF} achieves the best average \gls*{RMSE} it is likely still computationally more expensive for a \gls*{PF} method with a sufficiently high number of particles (e.g., it seems that 50 to 60 particles may suffice to achieve the same \gls*{RMSE} but with half the run time). 
We hypothesis that the estimation accuracy and computational runtime will be more favorable for the new method in future testing of higher dimensional models. 

To further validate that the trends and observations hold generally for the application of the methods to the stochastic \gls*{L63} model, we perform another \gls*{MC} simulation, but with 100 runs for each of the 10 initial means given in Table \ref{tab:mc_ic_sweep_ess_rmse} and labeled in Fig. \ref{fig:l63_ic_locations}. 
The \gls*{PF} exhibits a consistently low \gls*{nESS} and relatively large \gls*{RMSE}, indicating significant weight degeneracy. 
The \gls*{nPF} improves the estimation accuracy in several cases, but does not reduce the estimation error across all initial conditions. 
Notably, the new \gls*{Var-nPF} approach achieves consistently lower \gls*{RMSE}, demonstrating improved robustness with respect to variations in the initial state of the \gls*{L63}. 
These results indicate that incorporating the modified variational approach enhances the accuracy of the filtering process. 
The computational runtime trends from Table \ref{tab:mc_summary_with_pf3} are maintained in this \gls*{MC} experiment, as shown by the last row of Table \ref{tab:mc_ic_sweep_ess_rmse}. 

\section{Conclusions}
\label{section: conclusions}

In this paper, we have introduced a variational pseudo-observation guided \gls*{nPF} for filtering nonlinear chaotic systems with sparse observations that may exhibit ``rare" events. 
By generating intermediate pseudo-observations within each observation interval, the proposed approach replaces long time-horizon \glspl*{OCP} with shorter ones. 
Numerical results on the stochastic \gls*{L63} show that the \gls*{Var-nPF} improves estimation accuracy relative to both \gls*{PF} and \gls*{nPF} while reducing computational cost. 
This provides gentler nudging of the particles in each subinterval, 
enabling more consistent control and promoting better weight diversity.

The observation interval used in this work was shorter than the average switching time between lobes of the \gls*{L63} attractor.
To better mimic a ``rare" event, future work will test on longer duration regimes that are more  
comparable to the switching timescale.
This setting will better test the sensitivity of the variational component of the \gls*{Var-nPF}. 

Future efforts will also investigate the use of regularized variational methods \cite{Tremolet2006}, 
as well as 
test the \gls*{Var-nPF} on higher-dimensional systems such as the Lorenz 1996 model \cite{Lorenz1996} and geophysical models demonstrating observed rare events \cite{Qiu:2000.jpo.30.8}. 


\bibliography{fusion_2026_var_nPF,additional_references}

\end{document}